# Fully Coupled Electromechanical Elastodynamic Model for Guided Wave Propagation Analysis

LUKE BORKOWSKI, KUANG LIU, and ADITI CHATTOPADHYAY
*School for Engineering of Matter, Transport, and Energy*
*Arizona State University, Tempe, AZ, 85287*

## ABSTRACT

Physics-based computational models play a key role in the study of wave propagation for structural health monitoring (SHM) and the development of improved damage detection methodologies. Due to the complex nature of guided waves, accurate and efficient computation tools are necessary to investigate the mechanisms responsible for dispersion, coupling, and interaction with damage. In this paper, a fully coupled electromechanical elastodynamic model for wave propagation in a heterogeneous, anisotropic material system is developed. The final framework provides the full three dimensional displacement and electrical potential fields for arbitrary plate and transducer geometries and excitation waveform and frequency. The model is validated theoretically and proven computationally efficient. Studies are performed with surface bonded piezoelectric sensors to gain insight into the physics of experimental techniques used for SHM. Collocated actuation of the fundamental Lamb wave modes is modeled over a range of frequencies to demonstrate mode tuning capabilities. The displacement of the sensing surface is compared to the piezoelectric sensor electric potential to investigate the relationship between plate displacement and sensor voltage output. Since many studies, including the ones investigated in this paper, are difficult to perform experimentally, the developed model provides a valuable tool for the improvement of SHM techniques.

## INTRODUCTION

Among the various techniques used for structural health monitoring (SHM) of aerospace, civil, and mechanical structures, guided wave-based techniques have been proven most effective because of their wide array of applications and sensitivity to multiple types of damage (Raghavan and Cesnik, 2007; Giurgiutiu, 2008; Andrews et al., 2008). One of the most promising guided wave-based approaches for damage detection in these structures is Lamb wave-based SHM techniques (Giurgiutiu, 2008; Alleyne and Cawley, 1993; Staszewski et al., 1997; Jha and Watkins, 2009), which typically involve exciting the structure with ultrasonic stress waves, collecting its structural response, and then processing this response for detection and *in-situ* characterization of damage. Lamb waves have the ability to travel long distances in plate-like structures; therefore, SHM techniques utilizing Lamb wave analysis have the potential to monitor large areas with few actuators and sensors (Giurgiutiu, 2008). The abundance of structures, in particular aerospace structural components, whose mechanical behavior resembles that of thin plates or shells, also contributes to the vast application of this technique. To excite the structure with a Lamb wave, piezoelectric transducers are often used due to their many advantages, particularly the ability to serve as both an actuator and sensor (Guo and Cawley, 1993; Diaz and Soutis, 2000; Giurgiutiu, 2008).

Lamb wave techniques have been used by many researchers for damage detection in both metallic and composite structures. However, most of these methods are data driven (Liu, Mohanty, and Chattopadhyay, 2010; Liu et al., 2011; Soni, Das, and Chattopadhyay, 2009). Conducting an experiment for every sensor location, wave form type and frequency, and damage type and severity can be time-consuming and expensive. The use of hybrid sensing approaches that combine experimental data with results from a physics-based model (referred to as *virtual sensing*) have been found to be more effective in damage detection of complex aerospace structures (Chattopadhyay, et al., 2009). These models provide insight into the damage mechanism, allowing for further optimization of SHM techniques.

The complexity of Lamb waves that are excited and sensed using piezoelectric actuators and sensors for SHM arises from their dispersive nature, existence of at least two modes at any given frequency of excitation, electromechanical coupling due to the piezoelectric phenomenon, interaction with damage and material discontinuities at various length scales, and the three-dimensional nature of the problem. It is advantageous, therefore, to have computational models to study the physics of wave propagation, which can aid in the development of accurate damage detection methodologies. Models for wave propagation also provide a means to interpret the results obtained from experiments since the full displacement, stress, and strain fields can be studied as opposed to only the sensor signal in the case of experiments.

Due to limitations associated with analytical wave propagation models for aerospace structures, such as the difficulty involved in modeling complex geometries and material architectures and accounting for damage, numerical models are employed to solve the elastodynamic wave equation for the desired geometry, boundary conditions, actuation signals, and material properties. Numerous numerical techniques exist for modeling elastic wave propagation, such as finite element method (Talbot and Przemieniecki, 1976; Zienkiewicz, 1989; Koshiba, Karakida, and Suzuki, 1984), finite strip elements (Cheung, 1976; Liu et al., 1995; Liu and Achenbach, 1995), boundary element method (Yamawaki and Saito, 1992; Cho and Rose, 1996), spectral element methods (Fornberg, 1998; Krawczuk and Ostachowicz, 2001), and local interaction simulation approach/sharp interface model (LISA/SIM) (Delsanto, 1992, 1994, 1997).

In materials with the presence of damage or other material discontinuities (Agostini et al., 2003; Lee and Staszewski, 2007), LISA/SIM has proven to be an effective, accurate, and computationally efficient modeling technique for wave propagation. One of the main advantages of LISA/SIM is its ability to model wave propagation across sharp material property interfaces without incurring significant numerical error caused by the smearing of material properties across cell interfaces (Delsanto, 1992, 1994, 1997). Lee and Staszewski (2007) modeled Lamb wave-based damage detection in metallic specimens using LISA/SIM. Sundararaman (2007) extended the technique to include adaptive grid spacing for higher spatial resolution in regions of geometric complexity. Most Lamb wave studies using this technique are carried out on 2D geometries for the reason of computational efficiency (Lee and Staszewski, 2007). Since Lamb waves only exist in 3D bounded media, the 2D models require the Lamb wave group and phase velocities to be provided *a priori* for the in-plane simulation while the wave propagation in the through-thickness direction is modeled separately. Modeling the 3D problem in this 2D fashion limits the usefulness and accuracy of the model. A full 3D model is required to account for the coupling between the separate Lamb wave modes and to represent the mode conversions and reflections caused by boundaries, damage, or other material discontinuities. While LISA/SIM has been proven to be an effective tool for guided wave-based SHM, Raghavan and Cesnik (2007) asserted that the application of this technique has been limited due to the lack of a fully coupled electromechanical elastodynamic formulation to account for Lamb wave excitation and sensing.

Modeling of guided wave (GW) excitation and sensing is crucial to wave propagation simulation techniques because of the complex coupling between the electrical excitation of the piezoelectric actuator, the subsequent mechanical response of the actuator and structure, and finally the mechanical and electrical response of the sensor. Previous work focused on modeling the excitation of guided waves has primarily been based on the theory of elasticity; it has utilized the 'plane-strain' assumption, and has been limited to 2D (Viktorov, 1967; Ditri and Rose, 1994). Extensions of the elasticity theory-based approaches to 3D have used *impulse point body force* (Santosa and Pao, 1989) and *generic surface point sources* (Wilcox, 2004) to model the GW excitation. Relatively little work, however, has been done on modeling structurally integrated piezoelectric actuators with finite dimensions. Moulin, Assaad, and Delebarre (2000) modeled a surface-mounted PZT using a coupled finite element method (FEM)-normal mode expansion method. Other researchers have also utilized the built-in piezoelectric elements in commercial finite element packages (Soni, Das, and Chattopadhyay, 2009) to model actuators and sensors for SHM applications. Mindlin plate theory incorporating transverse shear and rotary inertia effects was used by other researchers to model the GW excitation as causing bending moments along the actuator edge (Rose and Wang, 2004; Veidt, Liu, and Kitipornchai, 2001). One major disadvantage of using Mindlin plate theory is that it can only model approximately the zeroth order antisymmetric Lamb wave mode and is only valid at low frequencies where no additional higher order antisymmetric modes are excited in the plate.

Giurgiutiu (2003) modeled an infinitely wide piezoelectric transducer to study the excitation of Lamb waves in an isotropic plate. He solved for the displacement and strain fields by first reducing the 3D elasticity problem to 2D using the Fourier integral theorem and then coming to a solution through inversion using residue theory. Raghavan and Cesnik (2005) developed an analytical modeling technique using 3D elasticity theory and the Fourier integral theorem to model actuators and sensors of finite dimensions. This approach was validated experimentally and numerically for the cases under investigation in the paper. However, the assumption made in the formulation of the analytical approach introduced in Raghavan and Cesnik (2005) limits its application to Lamb wave analysis in infinite plates without considering the effect of the actuator and sensor on structural dynamics and wave behavior since the actuation is modeled as causing an in-plane traction of uniform magnitude only along its perimeter in the direction normal to the free edge of the plate surface. In addition, the plate through-thickness displacement is not provided with this approach.

In the current paper, a fully coupled electromechanical elastodynamic model for wave propagation in a heterogeneous, anisotropic material system is developed. The objective of developing this novel modeling scheme is to accurately and efficiently study the physics of GW propagation for the purpose of SHM, and, in turn, ease the monitoring strategy used for damage detection with guided waves. The final set of equations provides the full 3D displacement and electrical potential fields for arbitrary plate and transducer geometries and excitation waveform and frequency. The model framework is based on that developed by Delsanto (1992, 1994, 1997) for an orthotropic material, but is extended to include piezoelectric coupling and explicit consideration of the piezoelectric actuators and sensors for an anisotropic material system. The model is validated theoretically by comparing the simulated wave speed to that predicted by Lamb wave theory over a wide range of frequency-thickness products. Various studies, which are difficult to conduct experimentally, are investigated for the governing physics of GW analysis for SHM. These studies include investigating the effect of actuation types on sensor signals, relative sensor voltage of Lamb wave modes excited with collocated actuators, and the relationship between the displacement components below the piezoelectric sensor with the sensor voltage.

## 3D ELECTROMECHANICAL COUPLED ELASTODYNAMIC MODEL FRAMEWORK

This section outlines the derivation of a set of incremental equations for the solution of a 3D fully generalized, fully coupled electromechanical elastodynamic wave propagation model for a heterogeneous, anisotropic material system. The final set of equations will provide the evolution of the time-varying displacement and electric potential fields for an arbitrary geometry and actuation waveform. This formulation solves the mechanical equations of motions as an initial value problem and Maxwell's equation as a boundary value problem at each time step. Since some details of the derivation are not discussed in the current paper, the reader is referred to Delsanto (1997) for the general procedure for solving the uncoupled elastodynamic wave equation for an orthotropic material using LISA/SIM, which inspired this formulation.

In this approach, the spatial domain is discretized in the *x*, *y*, *z* directions into a cuboidal grid with dimensions $\Delta x$, $\Delta y$, and $\Delta z$, respectively, as shown in Figure 1. The material properties of each cell are defined at the lower left front corner of the cell, meaning an element with its center at location $(\alpha + \Delta x/2, \beta + \Delta y/2, \gamma + \Delta z/2)$ will have its mechanical and physical properties defined at $(\alpha, \beta, \gamma)$. While the material properties are constant within each cell, they are allowed to vary across cells. The incorporation of SIM into the LISA framework allows for the accurate simulation of wave propagation across sharp material boundaries since the material properties are not smeared across cell interfaces. Additional points are defined in the grid, denoted by a star and a cross in Figure 1, at infinitesimal distances $\delta$ and $\iota$ from the nodal points and the interface in order to enforce continuity of displacement at the nodes and traction across the interface. The distances $\delta$ and $\iota$ are exaggerated in Figure 1 for clarity.

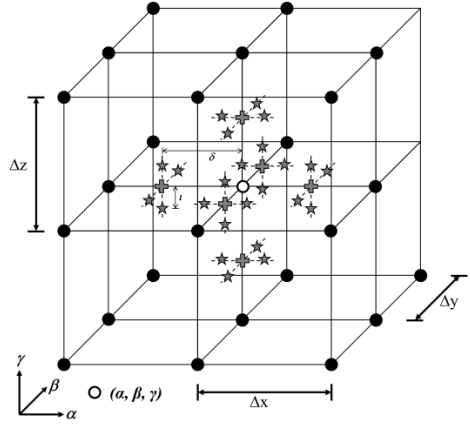

Figure 1: Definition of grid and supplemental points

For a linear elastic piezoelectric material, the constitutive equation that governs the interaction of the elastic and electric fields can be written as

$$\sigma_{ij} = C_{ijkl}\varepsilon_{kl} - e_{kij}E_k , \qquad \text{Equation (1)}$$

where $\sigma_{ij}$, $C_{ijkl}$, $\varepsilon_{kl}$, $e_{kij}$, and $E_k$ are the second order stress tensor, fourth order stiffness tensor, second order strain tensor, third order piezoelectric tensor, and first order electric field tensor, respectively. In addition, the electric displacement vector can be expressed in terms of the strain and electric field in the form

$$D_i = e_{ijk}\varepsilon_{jk} + \kappa_{ij}E_j , \qquad \text{Equation (2)}$$

where $D_i$ is the first order electric displacement tensor and $\kappa_{ij}$ is the second order dielectric tensor.

The components of the small strain tensor $\varepsilon_{kl}$ are expressed in terms of the displacement components $u_k$ using the strain-displacement relation,

$$\varepsilon_{kl} = \frac{1}{2}\left(u_{k,l} + u_{l,k}\right), \qquad \text{Equation (3)}$$

and the components of the electric field $E_i$ are obtained from the electric potential $\phi_i$ via

$$E_i = -\phi_{,i} . \qquad \text{Equation (4)}$$

Using the strain-displacement relation (Equation (3)), definition of electric field (Equation (4)), and the symmetry of the stiffness tensor, Equation (1) and Equation (2) can be expressed in terms of displacement and electric potential as

$$\sigma_{ij} = C_{ijkl}u_{k,l} + e_{kij}\phi_{,k} \qquad \text{Equation (5)}$$

and

$$D_i = e_{ijk}u_{j,k} - \kappa_{ij}\phi_{,j} \qquad \text{Equation (6)}$$

In an elastic medium, force equilibrium is enforced through the elastodynamic wave equation in the form

$$C_{ijkl}u_{k,jl} + e_{kij}\phi_{,kj} = \rho\ddot{u}_i . \qquad \text{Equation (7)}$$

Viscoelasticity was not included in the current paper since this has been previously investigated by previous researchers, such as Sundararaman (2007).

In the absence of volume charge, Maxwell's equation,

$$\nabla \cdot \mathbf{D} = 0 , \qquad \text{Equation (8)}$$

must be satisfied, which requires

$$e_{ijk}u_{j,ki} - \kappa_{ij}\phi_{,ji} = 0 . \qquad \text{Equation (9)}$$

A central difference scheme is used to approximate the second order derivatives of the displacement and electrical potential at points defined at $(\alpha + a\delta, \beta + b\delta, \gamma + c\delta)$ in the cuboidal grid in terms of their first order derivatives. Here, a, b, c represent neighboring nodes and have the value of ±1, ±1, ±1 and $\delta$ represents a small distance away from the node. Some of the expressions for the second order differential equations are supplied here for clarity.

$$u_{k,11}^{\alpha+a\delta,\beta+b\delta,\gamma+c\delta} = \frac{u_{k,1}^{\alpha+a/2,\beta,\gamma} - u_{k,1}^{\alpha+a\delta,\beta+b\delta,\gamma+c\delta}}{a\Delta x / 2} \qquad \text{Equation (10)}$$

$$u_{k,12}^{\alpha+a\delta,\beta+b\delta,\gamma+c\delta} = \frac{u_{k,2}^{\alpha+a,\beta+b/2,\gamma} - u_{k,2}^{\alpha,\beta+b/2,\gamma}}{a\Delta x} \qquad \text{Equation (11)}$$

$$\phi_{,11}^{\alpha+a\delta,\beta+b\delta,\gamma+c\delta} = \frac{\phi_{,1}^{\alpha+a/2,\beta,\gamma} - \phi_{,1}^{\alpha+a\delta,\beta+b\delta,\gamma+c\delta}}{a\Delta x / 2} \qquad \text{Equation (12)}$$

$$\phi_{,12}^{\alpha+a\delta,\beta+b\delta,\gamma+c\delta} = \frac{\phi_{,2}^{\alpha+a,\beta+b/2,\gamma} - \phi_{,2}^{\alpha,\beta+b/2,\gamma}}{a\Delta x} \qquad \text{Equation (13)}$$

Similarly, the first order derivative of displacement and electric potential at points $(\alpha+a/2,\beta,\gamma)$, $(\alpha,\beta+b/2,\gamma)$, and $(\alpha,\beta,\gamma+c/2)$ are also expressed using finite difference. For brevity, these expressions are not included in the present paper.

Next continuity of displacement and electric potential will be enforced at additional points defined at a small distance from the grid points. A very small distance, denoted by $\iota$ will be defined as

$$\iota = \delta^x, x \gg 1. \qquad \text{Equation (14)}$$

Since the procedure for enforcing continuity of the first derivative of displacement is similar to that of Delsanto (1997), it is not repeated in this paper. However, the continuity of the first derivative of electric potential results in,

$$\phi_{,1}^{\alpha+a\iota,\beta+b\delta,\gamma+c\delta} = \phi_{,1}^{\alpha+a\delta,\beta+b\delta,\gamma+c\delta} \qquad \phi_{,1}^{\alpha+a\delta,\beta+b\iota,\gamma+c\delta} = \frac{\phi^{\alpha+a,\beta,\gamma} - \phi^{\alpha,\beta,\gamma}}{a\Delta x} \qquad \phi_{,1}^{\alpha+a\delta,\beta+b\delta,\gamma+c\iota} = \frac{\phi^{\alpha+a,\beta,\gamma} - \phi^{\alpha,\beta,\gamma}}{a\Delta x}$$

Equation (15)     Equation (16)     Equation (17)

$$\phi_{,2}^{\alpha+a\iota,\beta+b\delta,\gamma+c\delta} = \frac{\phi^{\alpha,\beta+b,\gamma} - \phi^{\alpha,\beta,\gamma}}{b\Delta y} \qquad \phi_{,2}^{\alpha+a\delta,\beta+b\iota,\gamma+c\delta} = \phi_{,2}^{\alpha+a\delta,\beta+b\delta,\gamma+c\delta} \qquad \phi_{,2}^{\alpha+a\delta,\beta+b\delta,\gamma+c\iota} = \frac{\phi^{\alpha,\beta+b,\gamma} - \phi^{\alpha,\beta,\gamma}}{b\Delta y}$$

Equation (18)     Equation (19)     Equation (20)

$$\phi_{,3}^{\alpha+a\iota,\beta+b\delta,\gamma+c\delta} = \frac{\phi^{\alpha,\beta,\gamma+c} - \phi^{\alpha,\beta,\gamma}}{c\Delta z} \qquad \phi_{,3}^{\alpha+a\delta,\beta+b\iota,\gamma+c\delta} = \frac{\phi^{\alpha,\beta,\gamma+c} - \phi^{\alpha,\beta,\gamma}}{c\Delta z} \qquad \phi_{,3}^{\alpha+a\delta,\beta+b\delta,\gamma+c\iota} = \phi_{,3}^{\alpha+a\delta,\beta+b\delta,\gamma+c\delta}$$

Equation (21)     Equation (22)     Equation (23)

The expressions for the first order derivatives in Equation (15), Equation (19), and Equation (23), in addition to their displacement counterparts, remain unknown. To solve for equilibrium and Maxwell's equation, continuity of tractions and electric displacement across the element interfaces are enforced. This will allow for the unknown first order derivatives to be eliminated.

Evaluating the elastodynamic equilibrium at the points $(\alpha+a\delta, \beta+b\delta, \gamma+c\delta)$ can be expressed as

$$C_{ijkl}^{\alpha+a\delta,\beta+b\delta,\gamma+c\delta} u_{k,lj}^{\alpha+a\delta,\beta+b\delta,\gamma+c\delta} + e_{lij}^{\alpha+a\delta,\beta+b\delta,\gamma+c\delta} \phi_{,lj}^{\alpha+a\delta,\beta+b\delta,\gamma+c\delta} = \rho^{\alpha+a\delta,\beta+b\delta,\gamma+c\delta} \ddot{u}_i^{\alpha+a\delta,\beta+b\delta,\gamma+c\delta} \qquad \text{Equation (24)}$$

for $a,b,c = \pm 1$.

The stress tensor at points near the nodes can be expressed as

$$\sigma_{ij}^{\alpha+a\delta,\beta+b\delta,\gamma+c\delta} = C_{ijkl}^{\alpha+a\delta,\beta+b\delta,\gamma+c\delta} u_{k,l}^{\alpha+a\delta,\beta+b\delta,\gamma+c\delta} + e_{lij}^{\alpha+a\delta,\beta+b\delta,\gamma+c\delta} \phi_{,l}^{\alpha+a\delta,\beta+b\delta,\gamma+c\delta} \qquad \text{Equation (25)}$$

for $a,b,c = \pm 1$.

Next, traction continuity is imposed across the cell interfaces at points near the nodes while recalling that the material properties (i.e., stiffness tensor, density, piezoelectric tensor, and dielectric tensor) are constant in each cell, for example,

$$C_{ijkl}^{\alpha+\iota,\beta+\delta,\gamma+\delta} = C_{ijkl}^{\alpha+\delta,\beta+\iota,\gamma+\delta} = C_{ijkl}^{\alpha+\delta,\beta+\delta,\gamma+\iota} = C_{ijkl}^{\alpha+\delta,\beta+\delta,\gamma+\delta} = C_{ijkl}^{\alpha,\beta,\gamma}. \qquad \text{Equation (26)}$$

Since the cell faces are orthogonal and aligned, the tractions can be expressed directly as the stress tensor. The vector equations can be expressed in compacted form as

$$\sigma_{i1}^{\alpha-\iota,\beta+b\delta,\gamma+c\delta} = \sigma_{i1}^{\alpha+\iota,\beta+b\delta,\gamma+c\delta} \qquad \text{Equation}$$

$$\sigma_{i2}^{\alpha+a\delta,\beta-\iota,\gamma+c\delta} = \sigma_{i2}^{\alpha+a\delta,\beta+\iota,\gamma+c\delta}$$ Equation (27)

$$\sigma_{i3}^{\alpha+a\delta,\beta+b\delta,\gamma-\iota} = \sigma_{i3}^{\alpha+\delta,\beta+b\delta,\gamma+\iota}$$ Equation (28)

Equation (29)

for a,b,c= ±1.

After substituting the expressions for stress into Equation (27), Equation (28), and Equation (29), replacing the first and second order spatial derivatives with their respective finite difference expressions in Equation (24) and Equation (25), and summing over a, b, and c, the unevaluated first order derivatives can be eliminated through a linear combination of the traction continuity and equilibrium equations. The time derivatives of the displacement are then expanded using finite difference, and the final expression for the nodal displacement at time t+Δt is achieved, as presented in Equation (30), Equation (32), Equation (33), and Equation (34). The solution of displacement at any point at time t+Δt, solved using forward integration, is a function of the material properties of the surrounding elements and the displacement and electric potential of the surrounding nodes at time t and t-Δt.

$$u_i^{\alpha,\beta,\gamma,t+1} = 2u_i^{\alpha,\beta,\gamma,t} - u_i^{\alpha,\beta,\gamma,t-1} + \frac{\delta t^2}{8\bar{\rho}} \sum_{a,b,c=\pm 1} (f+g+h)$$ Equation (30)

where

$$\bar{\rho} = \frac{1}{8} \sum_{a,b,c=\pm 1} \rho^s,$$ Equation (31)

and

$$f = 2\left(\frac{f_x}{\Delta x^2} + \frac{f_y}{\Delta y^2} + \frac{f_z}{\Delta z^2}\right)$$

$$f_x = C_{i1k1}^s \left(u_k^{\alpha+a,\beta,\gamma} - u_k^{\alpha,\beta,\gamma}\right) + e_{1i1}^s \left(\phi^{\alpha+a,\beta,\gamma} - \phi^{\alpha,\beta,\gamma}\right)$$

$$f_y = C_{i2k2}^s \left(u_k^{\alpha,\beta+b,\gamma} - u_k^{\alpha,\beta,\gamma}\right) + e_{2i2}^s \left(\phi^{\alpha,\beta+b,\gamma} - \phi^{\alpha,\beta,\gamma}\right)$$

$$f_z = C_{i3k3}^s \left(u_k^{\alpha,\beta,\gamma+c} - u_k^{\alpha,\beta,\gamma}\right) + e_{3i3}^s \left(\phi^{\alpha,\beta,\gamma+c} - \phi^{\alpha,\beta,\gamma}\right),$$

Equation (32)

and

$$g = 2\left(\frac{g_{xy}}{ab\Delta x\Delta y} + \frac{g_{xz}}{ac\Delta x\Delta z} + \frac{g_{yx}}{ab\Delta x\Delta y} + \frac{g_{yz}}{bc\Delta y\Delta z} + \frac{g_{zx}}{ac\Delta x\Delta z} + \frac{g_{zy}}{bc\Delta y\Delta z}\right)$$

$$g_{xy} = C_{i1k2}^s \left(u_k^{\alpha,\beta+b,\gamma} - u_k^{\alpha,\beta,\gamma}\right) + e_{2i1}^s \left(\phi^{\alpha,\beta+b,\gamma} - \phi^{\alpha,\beta,\gamma}\right)$$

$$g_{xz} = C_{i1k3}^s \left(u_k^{\alpha,\beta,\gamma+c} - u_k^{\alpha,\beta,\gamma}\right) + e_{3i1}^s \left(\phi^{\alpha,\beta,\gamma+c} - \phi^{\alpha,\beta,\gamma}\right)$$

$$g_{yx} = C_{i2k1}^s \left(u_k^{\alpha+a,\beta,\gamma} - u_k^{\alpha,\beta,\gamma}\right) + e_{1i2}^s \left(\phi^{\alpha+a,\beta,\gamma} - \phi^{\alpha,\beta,\gamma}\right)$$

$$g_{yz} = C_{i2k3}^s \left(u_k^{\alpha,\beta,\gamma+c} - u_k^{\alpha,\beta,\gamma}\right) + e_{3i2}^s \left(\phi^{\alpha,\beta,\gamma+c} - \phi^{\alpha,\beta,\gamma}\right)$$

$$g_{zx} = C_{i3k1}^s \left(u_k^{\alpha+a,\beta,\gamma} - u_k^{\alpha,\beta,\gamma}\right) + e_{1i3}^s \left(\phi^{\alpha+a,\beta,\gamma} - \phi^{\alpha,\beta,\gamma}\right)$$

$$g_{zy} = C_{i3k2}^s \left(u_k^{\alpha,\beta+b,\gamma} - u_k^{\alpha,\beta,\gamma}\right) + e_{2i3}^s \left(\phi^{\alpha,\beta+b,\gamma} - \phi^{\alpha,\beta,\gamma}\right),$$

Equation (33)

and

$$h = \frac{h_{xy}}{ab\Delta x\Delta y} + \frac{h_{xz}}{ac\Delta x\Delta z} + \frac{h_{yz}}{bc\Delta y\Delta z}$$

$$h_{xy} = \left(C^s_{i1k2} + C^s_{i2k1}\right)\left(u_k^{\alpha+a,\beta+b,\gamma} - u_k^{\alpha+a,\beta,\gamma} - u_k^{\alpha,\beta+b,\gamma} + u_k^{\alpha,\beta,\gamma}\right) +$$
$$\left(e^s_{1i2} + e^s_{2i1}\right)\left(\phi^{\alpha+a,\beta+b,\gamma} - \phi^{\alpha+a,\beta,\gamma} - \phi^{\alpha,\beta+b,\gamma} + \phi^{\alpha,\beta,\gamma}\right)$$

$$h_{xz} = \left(C^s_{i1k3} + C^s_{i3k1}\right)\left(u_k^{\alpha+a,\beta,\gamma+c} - u_k^{\alpha+a,\beta,\gamma} - u_k^{\alpha,\beta,\gamma+c} + u_k^{\alpha,\beta,\gamma}\right) +$$
$$\left(e^s_{1i3} + e^s_{3i1}\right)\left(\phi^{\alpha+a,\beta,\gamma+c} - \phi^{\alpha+a,\beta,\gamma} - \phi^{\alpha,\beta,\gamma+c} + \phi^{\alpha,\beta,\gamma}\right)$$

$$h_{yz} = \left(C^s_{i2k3} + C^s_{i3k2}\right)\left(u_k^{\alpha,\beta+b,\gamma+c} - u_k^{\alpha,\beta+b,\gamma} - u_k^{\alpha,\beta,\gamma+c} + u_k^{\alpha,\beta,\gamma}\right) +$$
$$\left(e^s_{2i3} + e^s_{3i2}\right)\left(\phi^{\alpha,\beta+b,\gamma+c} - \phi^{\alpha,\beta+b,\gamma} - \phi^{\alpha,\beta,\gamma+c} + \phi^{\alpha,\beta,\gamma}\right)$$

Equation (34)

where superscript "s" denotes the point $(\alpha+a\delta,\beta+b\delta,\gamma+c\delta)$.

A similar approach is followed to achieve an expression for the electric potential at time t. First, Maxwell's equation is enforced at every point $(\alpha+a\delta,\beta+b\delta,\gamma+c\delta)$ as

$$e_{ijk}^{\alpha+a\delta,\beta+b\delta,\gamma+c\delta} u_{j,ki}^{\alpha+a\delta,\beta+b\delta,\gamma+c\delta} - \kappa_{ij}^{\alpha+a\delta,\beta+b\delta,\gamma+c\delta} \phi_{,ji}^{\alpha+a\delta,\beta+b\delta,\gamma+c\delta} = 0$$

Equation (35)

for a,b,c= ±1.

Next the continuity of the normal electric displacements are enforced at infinitesimal distances from the interface, which will result in the following equations,

$$D_1^{\alpha+\iota,\beta+b\delta,\gamma+c\delta} = D_1^{\alpha-\iota,\beta+b\delta,\gamma+c\delta}$$

Equation (36)

$$D_2^{\alpha+a\delta,\beta+\iota,\gamma+c\delta} = D_2^{\alpha+a\delta,\beta-\iota,\gamma+c\delta}$$

Equation (37)

$$D_3^{\alpha+a\delta,\beta+b\delta,\gamma+\iota} = D_3^{\alpha+a\delta,\beta+b\delta,\gamma-\iota}$$

Equation (38)

for a,b,c= ±1.

After substituting the expressions for electric displacement (Equation (6)) into Equation (36), Equation (37), and Equation (38)), replacing the first order spatial derivatives with their respective finite difference expressions in Equation (6) and Equation (9), and summing over a, b, and c, the unevaluated first order derivatives can be eliminated through a linear combination of the electric displacement continuity and Maxwell's equation. After simplification, the final expression for the electric potential at time t is achieved, as seen in Equation (39), Equation (40), Equation (41), and Equation (42). The solution of electric potential at any point at time t is a function of the material properties of the surrounding elements and the displacement and electric potential of the surrounding nodes at time t. Since the coupled equation for electric displacement at the point $(\alpha,\beta,\gamma)$ at the current time step is dependent on the electric potential of the nodes surrounding the point $(\alpha,\beta,\gamma)$ at the current time step, the solution of the boundary value problem requires a linear algebra technique for the solution of a set of dependent equations. For the current paper, LU decomposition was utilized to solve for the electric potential for the reasons of computational accuracy and efficiency.

$$\sum_{a,b,c=\pm 1}(q+r+s)=0$$

Equation (39)

where

$$q = 2\left(\frac{q_x}{\Delta x^2} + \frac{q_y}{\Delta y^2} + \frac{q_z}{\Delta z^2}\right)$$

$$q_x = e_{1j1}^s \left(u_j^{\alpha+a,\beta,\gamma} - u_j^{\alpha,\beta,\gamma}\right) - \kappa_{11}^s \left(\phi^{\alpha+a,\beta,\gamma} - \phi^{\alpha,\beta,\gamma}\right)$$

$$q_y = e_{2j2}^s \left(u_j^{\alpha,\beta+b,\gamma} - u_j^{\alpha,\beta,\gamma}\right) - \kappa_{22}^s \left(\phi^{\alpha,\beta+b,\gamma} - \phi^{\alpha,\beta,\gamma}\right)$$

$$q_z = e_{3j3}^s \left(u_j^{\alpha,\beta,\gamma+c} - u_j^{\alpha,\beta,\gamma}\right) - \kappa_{33}^s \left(\phi^{\alpha,\beta,\gamma+c} - \phi^{\alpha,\beta,\gamma}\right),$$

Equation (40)

and

$$r = 2\left(\frac{r_{xy}}{ab\Delta x \Delta y} + \frac{r_{xz}}{ac\Delta x \Delta z} + \frac{r_{yx}}{ab\Delta x \Delta y} + \frac{r_{yz}}{bc\Delta y \Delta z} + \frac{r_{zx}}{ac\Delta x \Delta z} + \frac{r_{zy}}{bc\Delta y \Delta z}\right)$$

$$r_{xy} = e_{1j2}^s \left(u_j^{\alpha,\beta+b,\gamma} - u_j^{\alpha,\beta,\gamma}\right) - \kappa_{12}^s \left(\phi^{\alpha,\beta+b,\gamma} - \phi^{\alpha,\beta,\gamma}\right)$$

$$r_{xz} = e_{1j3}^s \left(u_j^{\alpha,\beta,\gamma+c} - u_j^{\alpha,\beta,\gamma}\right) - \kappa_{13}^s \left(\phi^{\alpha,\beta,\gamma+c} - \phi^{\alpha,\beta,\gamma}\right)$$

$$r_{yx} = e_{2j1}^s \left(u_j^{\alpha+a,\beta,\gamma} - u_j^{\alpha,\beta,\gamma}\right) - \kappa_{21}^s \left(\phi^{\alpha+a,\beta,\gamma} - \phi^{\alpha,\beta,\gamma}\right)$$

$$r_{yz} = e_{2j3}^s \left(u_j^{\alpha,\beta,\gamma+c} - u_j^{\alpha,\beta,\gamma}\right) - \kappa_{23}^s \left(\phi^{\alpha,\beta,\gamma+c} - \phi^{\alpha,\beta,\gamma}\right)$$

$$r_{zx} = e_{3j1}^s \left(u_j^{\alpha+a,\beta,\gamma} - u_j^{\alpha,\beta,\gamma}\right) - \kappa_{31}^s \left(\phi^{\alpha+a,\beta,\gamma} - \phi^{\alpha,\beta,\gamma}\right)$$

$$r_{zy} = e_{3j2}^s \left(u_j^{\alpha,\beta+b,\gamma} - u_j^{\alpha,\beta,\gamma}\right) - \kappa_{32}^s \left(\phi^{\alpha,\beta+b,\gamma} - \phi^{\alpha,\beta,\gamma}\right),$$

Equation (41)

and

$$s = \frac{s_{xy}}{ab\Delta x \Delta y} + \frac{s_{xz}}{ac\Delta x \Delta z} + \frac{s_{yz}}{bc\Delta y \Delta z}$$

$$s_{xy} = \left(e_{1j2}^s + e_{2j1}^s\right)\left(u_j^{\alpha+a,\beta+b,\gamma} - u_j^{\alpha+a,\beta,\gamma} - u_j^{\alpha,\beta+b,\gamma} + u_j^{\alpha,\beta,\gamma}\right) - \left(\kappa_{12}^s + \kappa_{21}^s\right)\left(\phi^{\alpha+a,\beta+b,\gamma} - \phi^{\alpha+a,\beta,\gamma} - \phi^{\alpha,\beta+b,\gamma} + \phi^{\alpha,\beta,\gamma}\right)$$

$$s_{xz} = \left(e_{1j3}^s + e_{3j1}^s\right)\left(u_j^{\alpha+a,\beta,\gamma+c} - u_j^{\alpha+a,\beta,\gamma} - u_j^{\alpha,\beta,\gamma+c} + u_j^{\alpha,\beta,\gamma}\right) - \left(\kappa_{13}^s + \kappa_{31}^s\right)\left(\phi^{\alpha+a,\beta,\gamma+c} - \phi^{\alpha+a,\beta,\gamma} - \phi^{\alpha,\beta,\gamma+c} + \phi^{\alpha,\beta,\gamma}\right)$$

$$s_{yz} = \left(e_{2j3}^s + e_{3j2}^s\right)\left(u_j^{\alpha,\beta+b,\gamma+c} - u_j^{\alpha,\beta+b,\gamma} - u_j^{\alpha,\beta,\gamma+c} + u_j^{\alpha,\beta,\gamma}\right) - \left(\kappa_{23}^s + \kappa_{32}^s\right)\left(\phi^{\alpha,\beta+b,\gamma+c} - \phi^{\alpha,\beta+b,\gamma} - \phi^{\alpha,\beta,\gamma+c} + \phi^{\alpha,\beta,\gamma}\right),$$

Equation (42)

where superscript "s" denotes the point $(\alpha + a\delta, \beta + b\delta, \gamma + c\delta)$.

## SIMULATIONS RESULTS AND DISCUSSION

**Physical Model Development**

A 247 mm x 247 mm x 4 mm aluminum plate with collocated actuators and a single sensor were modeled for the studies presented in this paper. The actuators and sensors were centered on the plate and separated by a distance of 22 mm. The aluminum plate was modeled as a homogeneous, isotropic material with a density of 2780 kg/m$^3$, Young's modulus of 70 GPa, and Poisson's ratio of 0.3. The orthotropic material properties of the lead zirconate titanate (PZT) piezoelectric actuators and sensors are presented in Table 1. A 5 cycle cosine tone burst signal was used to excite the PZT actuators with a maximum electric potential of 10 V.

Table 1. PZT (APC 850) properties

| Elastic Properties | | | | | |
|---|---|---|---|---|---|
| **Elastic Moduli (Pa)** | | **Poisson's Ratio** | | **Shear Moduli (Pa)** | |
| **E1** | 6.30e10 | **n12** | 0.301 | **G12** | 2.35e10 |
| **E2** | 6.30e10 | **n13** | 0.532 | **G13** | 2.30e10 |
| **E3** | 5.40e10 | **n23** | 0.532 | **G23** | 2.30e10 |
| **Density (kg/m$^3$)** | 7500 | | | | |
| Piezoelectric Properties (C/m$^2$) | | | | | |
| **e1 11** | 0 | **e2 11** | 0 | **e3 11** | 2.18 |
| **e1 22** | 0 | **e2 22** | 0 | **e3 22** | 2.18 |
| **e1 33** | 0 | **e2 33** | 0 | **e3 33** | 23.59 |
| **e1 12** | 0 | **e2 12** | 0 | **e3 12** | 0 |
| **e1 13** | 27.14 | **e2 13** | 0 | **e3 13** | 0 |
| **e1 23** | 0 | **e2 23** | 27.14 | **e3 23** | 0 |
| Dielectric Properties (C/V-m) | | | | | |
| **κ11** | 1.51e-8 | **κ22** | 1.51e-8 | **κ33** | 1.30e-8 |

Issues that must be considered when implementing the current numerical framework for wave propagation modeling are convergence, numerical dispersion, and pulse and amplitude distortions. Several factors contribute to these issues. Pulse distortion, for example, can be mitigated by satisfying the Courant Friedrich Lewy (CFL) number, Equation (43).

$$CFL = c_{max} \Delta t \sqrt{\frac{1}{\Delta x^2} + \frac{1}{\Delta y^2} + \frac{1}{\Delta z^2}} \leq 1,$$  Equation (43)

where $c_{max}$ is the maximum wave speed (i.e., longitudinal wave speed), $\Delta t$ is the time step (i.e., sampling period), and $\Delta x$, $\Delta y$, $\Delta z$ are the grid spacings for the cuboid elements (Virieux, 1986). To prevent amplitude distortion, the general criterion is to have at least eight elements per minimum wavelength (Balasubramanyam et al., 1996). It is also commonly advised to avoid having more than twenty elements per minimum wavelength to avoid computational issues such as long run times and numerical error associated with the propagation of round-off error (Alleyne and Cawley, 1991). The grid spacings (i.e., $\Delta x$, $\Delta y$, and $\Delta z$) and time step (i.e., $\Delta t$) for the studies presented in this paper were chosen to ensure convergence while minimizing numerical error and computational load. The grid spacings in the $\Delta x$, $\Delta y$, and $\Delta z$ directions were held at 1 mm while the time step was adjusted to satisfy the CFL criterion.

**Collocated Actuators for Selective Lamb Wave Mode Suppression**

Collocated piezoelectric actuators have been utilized experimentally to selectively suppress Lamb wave modes for the purpose of SHM (Sohn and Kim, 2010). The suppression of one of the two fundamental Lamb wave modes (symmetric or antisymmetric) is achieved by selectively poling the collocated piezoelectric actuators, indicated with the black arrows in Figure 2. In SHM it is often desired to excite a wave with predominately symmetric or antisymmetric behavior to facilitate time-of-arrival calculation or to tailor the Lamb wave excitation to the type and location of damage in the structure. Although this phenomenon has been proven theoretically (Sohn and Kim, 2010), it is difficult to replicate experimentally. Slight variance in the relative actuator placement or in the piezoelectric actuator properties can have a significant impact on the degree of mode suppression. Numerical wave propagation models offer a valuable tool in investigating the physics of this experimental technique. The ability to separately model the fundamental Lamb wave modes is necessary for understanding the role each mode plays in the overall propagation of the Lamb wave and its interaction with damage and other features.

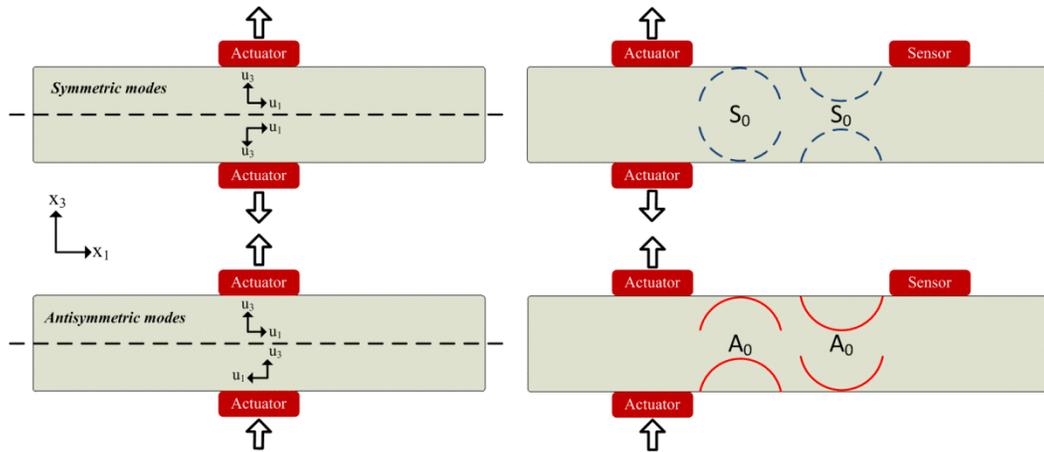

Figure 2. Relative actuator voltage poling directions and resultant through-thickness displacement profile for collocated piezoelectric actuation for selective Lamb wave mode suppression

Giurgiutiu (2005) demonstrated the concept of Lamb wave mode tuning using a single actuator, which involves exciting the structure with a frequency at which the $A_0$ or $S_0$ mode is most prevalent (i.e., largest relative energy). This type of selective tuning is possible because the energy of each mode varies with frequency. Although mode tuning using a single piezoelectric actuator has proven to be feasible, larger suppression of the undesired mode can be achieved with collocated actuators, as seen in Figure 2. Given the difficulty in implementing this technique experimentally, numerical models can be called to investigate the physics and provide insight into the problem before experimental implementation occurs. For damage detection using Lamb waves, it is desirable to excite a mode with the largest possible energy. Since the use of collocated actuators cannot completely eliminate the undesired mode, it is beneficial to know the frequency at which the maximum difference between the mode energies occurs. For homogeneous, isotropic specimens, an analytical technique such as the one utilized by Giurgiutiu (2005) can predict the relative mode energies; however, for specimens with complex heterogeneous architectures and anisotropy, numerical models such as the one presented in the current paper are required.

To demonstrate the concept, a 4 mm thick aluminum plate was modeled with collocated actuators and a single sensor. Through modeling collocated actuators to selectively excite a single mode at various frequencies and by overlaying the sensor voltage results, the relative sensor energies can be compared, as seen in Figure 3a, Figure 3b, Figure 3c, and Figure 3d for the frequency-half thickness products (fb/2) of 200 kHz-mm, 300 kHz-mm, 400 kHz-mm, and 500 kHz-mm, respectively.

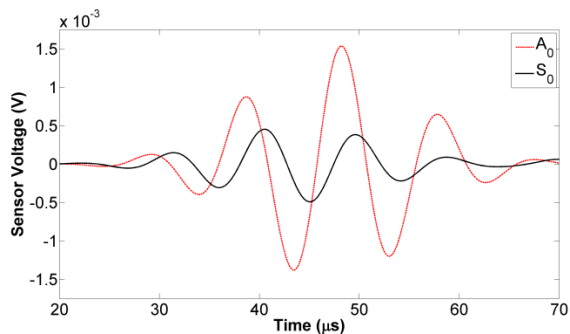
a) Sensor voltage vs. time for fb/2=200 kHz-mm

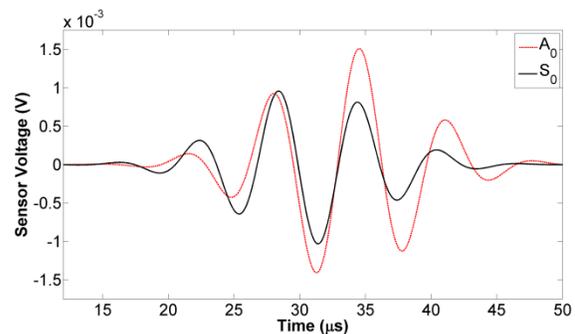
b) Sensor voltage vs. time for fb/2=300 kHz-mm

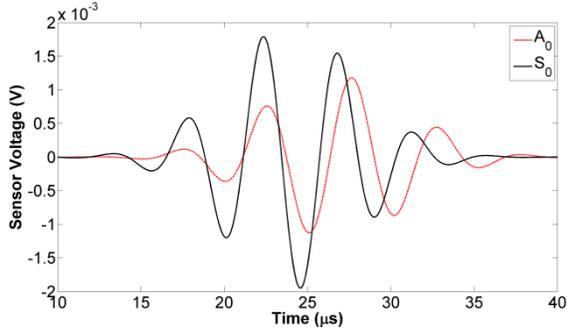
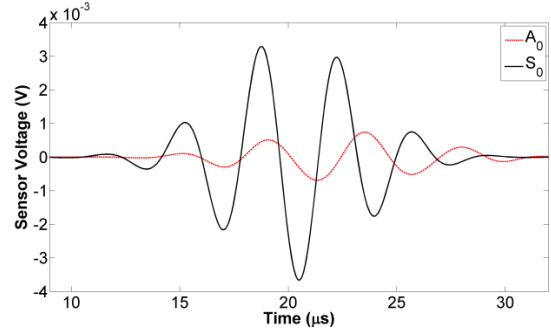

    c) Sensor voltage vs. time for fb/2=400 kHz-mm      d) Sensor voltage vs. time for fb/2=500 kHz-mm

Figure 3. Comparison of sensor voltage between symmetric and antisymmetric zeroth order Lamb wave modes for fb/2 equal to a) 200 kHz-mm, b) 300 kHz-mm, c) 400 kHz-mm, and d) 500 kHz-mm

**Effect of Actuation Type**

    Before development of the fully coupled electromechanical theory for LISA/SIM, researchers wishing to model piezoelectric actuation (Lee and Staszewski, 2007; Sundararaman, 2007) were forced to apply displacements to the 'piezoelectric' nodes or the cells beneath the actuator. In a study on the excitation of surface-bonded piezoelectric transducers, Giurgiutiu (2003) noted that the actuators typically used for SHM operated in a 'pinching' fashion or by causing a traction tangent to the plate surface. Most researchers found that application of an actuation in the form of a displacement in the in-plane direction gave more accurate prediction of wave speeds. However, this type of actuation does not take into consideration the complex piezoelectric coupling occurring within the actuator that causes the application of traction to the plate surface as a result of the externally supplied voltage across the actuator. To justify representing the piezoelectric actuation with applied displacement, an investigation into the effects is necessary.

    A study was conducted to investigate the effect of and error incurred due to displacement actuation compared to explicitly modeling the piezoelectric device. Three commonly used actuation types were investigated: electrical actuation, displacement in the y-direction actuation, and displacement in the z-direction actuation, as shown in Figure 4. The sensor signals for the $A_0$ and $S_0$ Lamb wave modes received from three different actuations are shown in Figure 5a and Figure 5b, respectively. It is evident from the plots that the time-of-arrival and wave speed of the displacement actuations vary significantly from that of the electrical actuation. Table 2 presents a comparison of the simulated $A_0$ and $S_0$ wave speeds ($v_s$) for each of the three actuation types compared to the theoretical wave speed ($v_t$) and the corresponding error. The theoretical wave speed was obtained by numerically solving the characteristic Lamb wave equations for the wave group velocity.

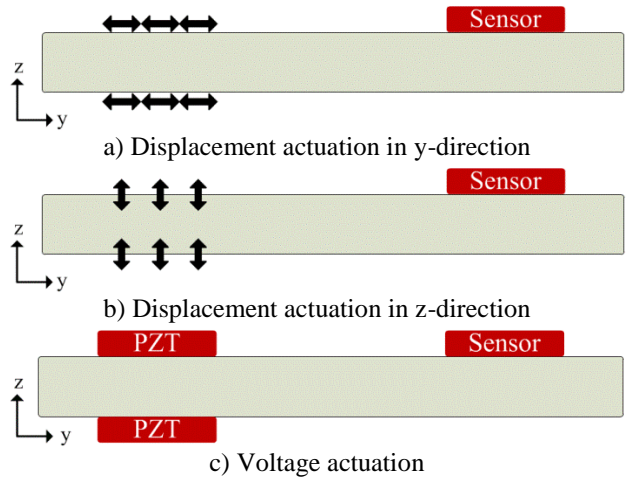

a) Displacement actuation in y-direction

b) Displacement actuation in z-direction

c) Voltage actuation

Figure 4. Excitation of GW in plate for three different actuation types: (a) displacement in the y-direction, (b) displacement in the z-direction, and (c) voltage actuation

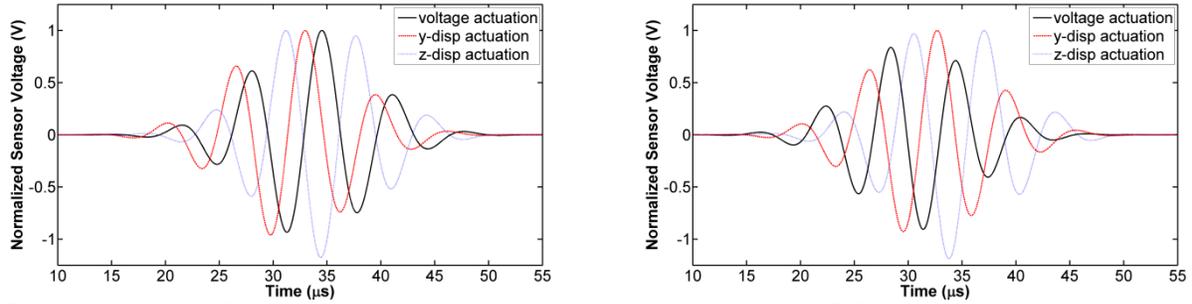

a) Sensor signal for $A_0$ Lamb wave mode

b) Sensor signal for $S_0$ Lamb wave mode

Figure 5. Sensor signal comparison for three different actuation types

Table 2. Comparison of simulated wave speeds using different actuation types

| Actuation | $v_s$ (m/s) | $v_t$ (m/s) | error |
|---|---|---|---|
| **$A_0$ mode** | | | |
| electrical | 2846.42 | 2965.73 | 4.02% |
| displacement (y) | 3598.30 | 2965.73 | -21.33% |
| displacement (z) | 2918.16 | 2965.73 | 1.60% |
| **$S_0$ mode** | | | |
| electrical | 4995.46 | 5192.83 | 3.80% |
| displacement (y) | 3774.23 | 5192.83 | 27.32% |
| displacement (z) | 3156.84 | 5192.83 | 39.21% |

Analysis of the data in Table 2 reveals inconsistency in wave speed that results from modeling the piezoelectric actuation as a displacement boundary condition. In particular, although the modeled wave speed for the z-direction displacement actuation is able to match the theoretical $A_0$ wave speed within 1.60%, its simulated $S_0$ wave speed is 39.21% below the theoretical wave speed. Actuating the plate with a y-direction displacement results in a simulated wave speed that is greater than 20% above the theoretical $A_0$ wave speed and greater than 25% below the theoretical $S_0$ wave speed. Due to the complex electromechanical coupling that occurs within a piezoelectric element, approximating the resultant displacement as unidirectional will produce inaccurate and inconsistent model results.

**Relationship between piezoelectric sensor displacement and output voltage**

An advantage to having an accurate numerical tool to simulate the excitation, propagation, and sensing of GW allows for investigating phenomena that cannot be studied in an experimental environment. A study was conducted to compare the electric potential of a sensor to the displacements below the sensor. This type of study is nearly impossible to conduct experimentally. Piezoelectric sensors were used to detect the displacement on the surface of the plate due to the presence of a propagating wave. Since the voltage output of the sensor is a function of the displacement gradient on its bonded surface, comparison of the displacement components of the interfacial nodes and the sensor voltage can provide physical insight into the mechanisms governing piezoelectric sensing. In Figure 6, it is shown that the voltage of the sensor lags the displacement in the y-direction beneath the sensor. This result is expected since strain in the piezoelectric sensor will slightly lag the displacement of its interfacial nodes. The shear component of the Lamb wave (z-displacement) has a similar time lag as compared to the sensor voltage.

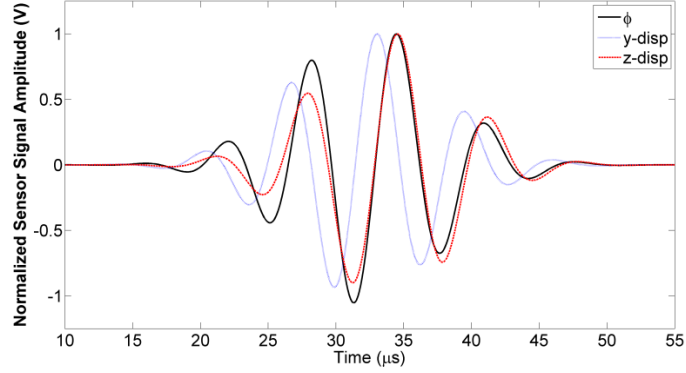

Figure 6. Comparison of sensor voltage and nodal displacement components beneath sensor

**Theoretical Validation**

The fully coupled electromechanical model was validated theoretically by modeling a 4 mm thick aluminum plate with collocated actuators and a single sensor, separated by a distance of 22 mm. The Lamb wave governing equations (Lamb, 1917) are shown in Equation (44) where the ±1 exponent indicates the symmetric and antisymmetric modes, respectively. The governing equations were solved numerically using the technique outlined in Rose (1999). The phase and group velocities can be then solved using Equation (46) and Equation (47).

$$\frac{\tan(\beta b/2)}{\tan(\alpha b/2)} = -\left\{\frac{4\alpha\beta k^2}{\left(k^2-\beta^2\right)^2}\right\}^{\pm 1} \qquad \text{Equation (44)}$$

where

$$\alpha^2 = \frac{\omega^2}{c_l^2} - k^2, \ \beta^2 = \frac{\omega^2}{c_t^2} - k^2, \qquad \text{Equation (45)}$$

and $\omega$ is the angular frequency, $k$ is the wave number, $c_l$ is the longitudinal wave speed, $c_t$ is the transverse wave speed, and $b$ is the plate thickness. The equations for the Lamb wave group and phase velocities are

$$c_p = \frac{\omega}{k} \text{ and} \qquad \text{Equation (46)}$$

$$c_g = \frac{d\omega}{dk}. \qquad \text{Equation (47)}$$

By utilizing collocated actuators, Lamb wave modes can be excited selectively, allowing for direct comparison between the simulated results and the theoretical dispersion curve for an aluminum plate over a range of frequencies commonly utilized for SHM. In addition to the commonly used frequencies, additional simulations were carried out to prove that the model can accurately predict the Lamb wave group velocity at higher frequency-thickness products as well. Giurgiutiu (2005) analytically determined that there is a limited frequency range in which the energy of the $S_0$ mode is greater than that of the $A_0$ mode. Because of this, collocated actuators are necessary for comparing the simulated $S_0$ group velocity to that predicted with Lamb wave theory. Since the time-of-arrival of the $A_0$ and $S_0$ Lamb wave modes is a common feature used in SHM damage detection methodologies, wave propagation models for this purpose must be able to accurately predict the wave speed of these zeroth order modes. Figure 7 presents a comparison between the simulated group velocities ($c_g$) vs. the frequency-half thickness product (fb/2) with the theoretical $A_0$ and $S_0$ group velocities. The discrepancies between the simulated and theoretical wave speed for some of the frequency-half thickness products investigated can be attributed to the time lag in the sensor voltage or from grid dispersion caused by insufficient spatial samples per wavelength.

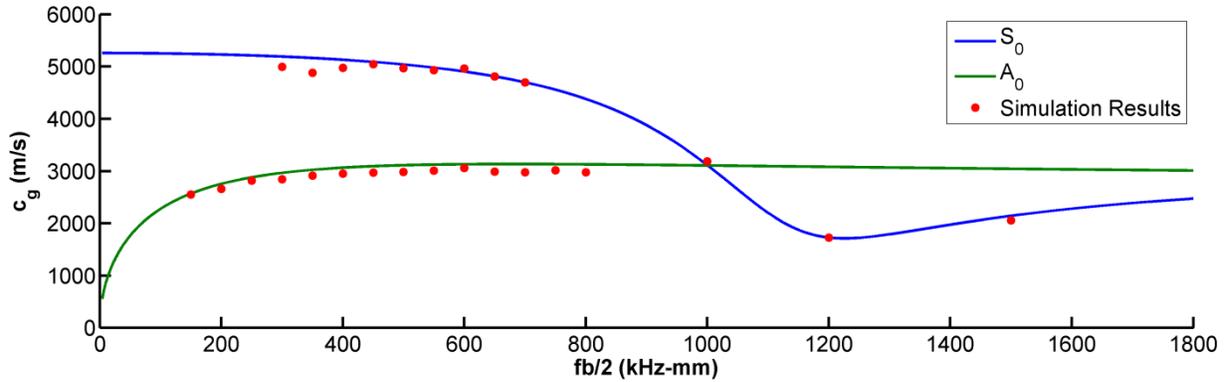

Figure 7. Theoretical validation of $A_0$ and $S_0$ Lamb mode group velocities

The two fundamental Lamb wave modes (symmetric and antisymmetric) have distinct displacement signatures in the plane orthogonal to the propagation direction of the wave. Lamb waves have the unique characteristic of resembling a standing wave through the thickness and a traveling wave in plane. Figure 2 demonstrates this phenomenon. For symmetric Lamb wave modes, the out-of-plane displacement profile is symmetric while the in-plane displacement is antisymmetric with respect to the mid-plane of the plate. For antisymmetric Lamb wave modes, the opposite is true. By plotting the out-of-plane and in-plane displacement and the displacement vectors at each node for the $A_0$ and $S_0$ modes, the characteristic displacement profiles of each mode can be visualized and compared as seen in Figure 8, Figure 9, Figure 11, and Figure 12 for two simulation times (16.625 μs and 33.25 μs). The out-of-plane displacement on the plate surface is shown in Figure 10 and Figure 13 for the simulation times of 16.625 μs and 33.25 μs, respectively. It should be noted that the contours are not to scale as they have been rescaled to clearly demonstrate the through-thickness displacement profile of Lamb waves.

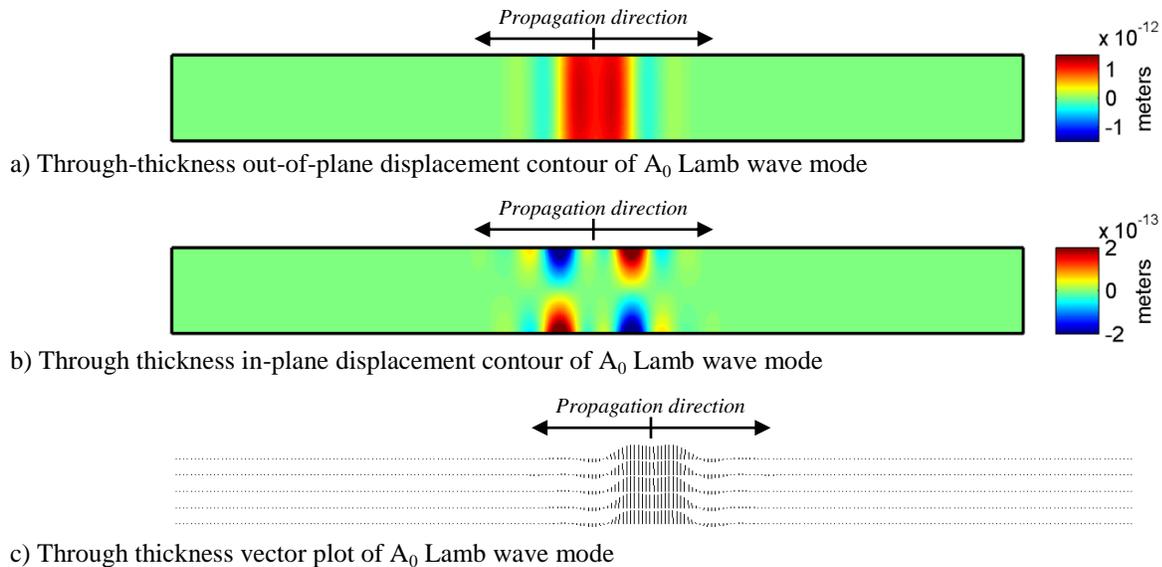

a) Through-thickness out-of-plane displacement contour of $A_0$ Lamb wave mode

b) Through thickness in-plane displacement contour of $A_0$ Lamb wave mode

c) Through thickness vector plot of $A_0$ Lamb wave mode

Figure 8. Through-thickness plots of (a) out-of-plane displacement, (b) in-plane displacement, and (c) vector field for $A_0$ Lamb wave mode at t=16.625 μs for fb/2=300 kHz-mm

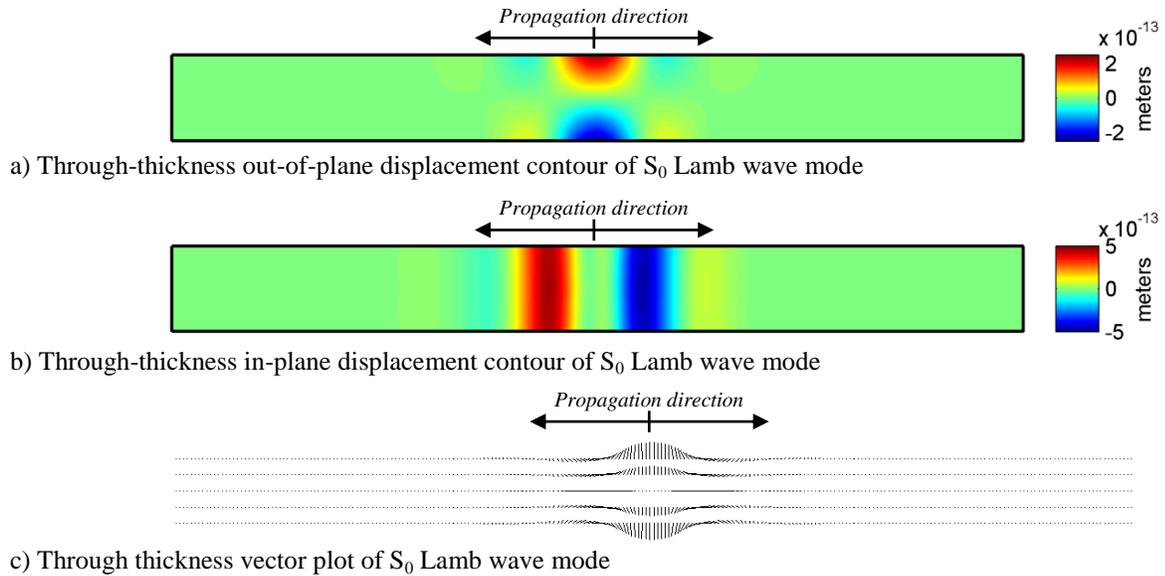

a) Through-thickness out-of-plane displacement contour of $S_0$ Lamb wave mode

b) Through-thickness in-plane displacement contour of $S_0$ Lamb wave mode

c) Through thickness vector plot of $S_0$ Lamb wave mode

Figure 9. Through-thickness plots of (a) out-of-plane displacement, (b) in-plane displacement, and (c) vector field for $S_0$ Lamb wave mode at t=16.625 μs for fb/2=300 kHz-mm

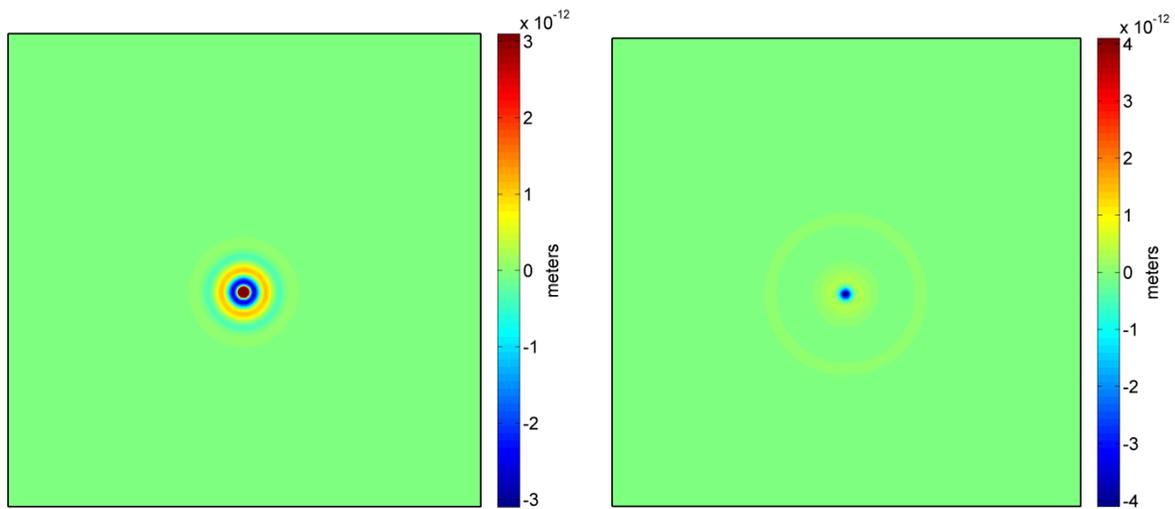

a) $A_0$ Lamb wave mode out-of-plane displacement       b) $S_0$ Lamb wave mode out-of-plane displacement

Figure 10. Out-of-plane displacement contour on the plate surface of the (a) $A_0$ and (b) $S_0$ Lamb wave modes at t=16.625 μs for fb/2=300 kHz-mm

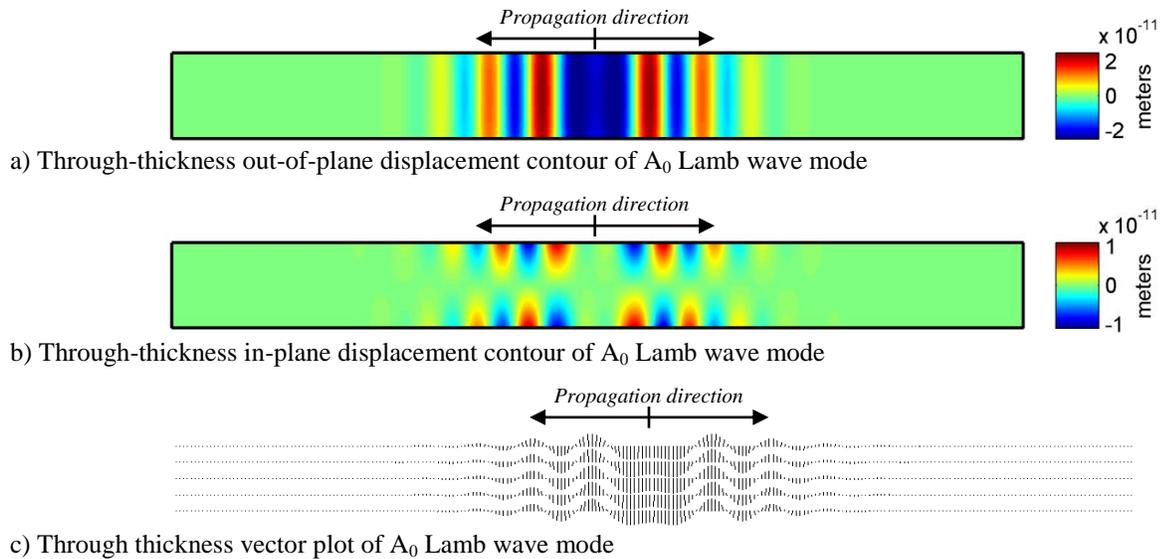

a) Through-thickness out-of-plane displacement contour of $A_0$ Lamb wave mode

b) Through-thickness in-plane displacement contour of $A_0$ Lamb wave mode

c) Through thickness vector plot of $A_0$ Lamb wave mode

Figure 11. Through-thickness plots of (a) out-of-plane displacement, (b) in-plane displacement, and (c) vector field for $A_0$ Lamb wave mode at t=33.25 μs for fb/2=300 kHz-mm

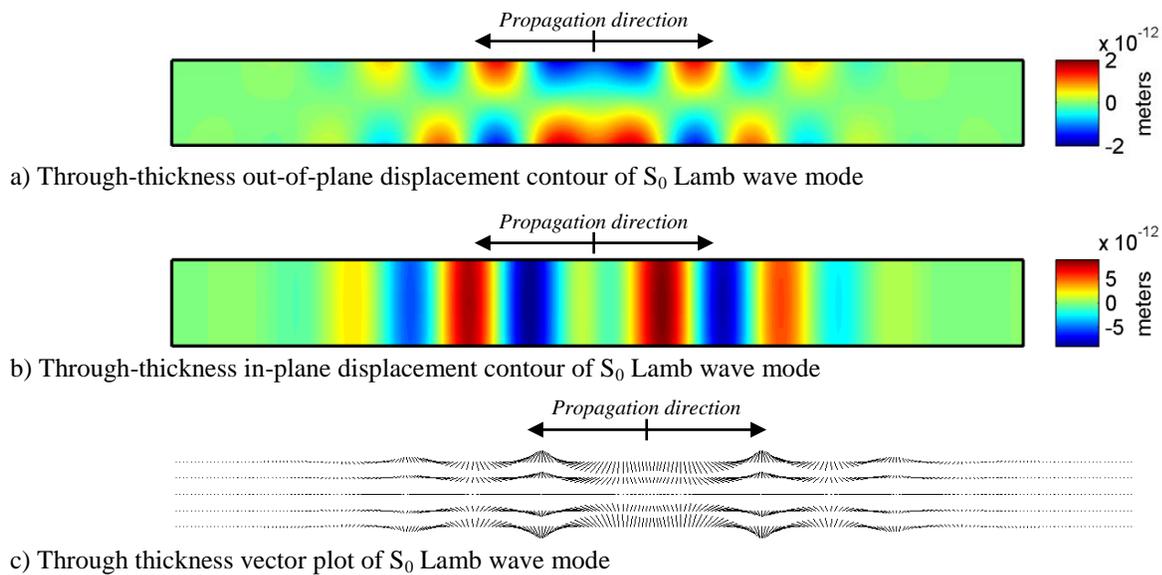

a) Through-thickness out-of-plane displacement contour of $S_0$ Lamb wave mode

b) Through-thickness in-plane displacement contour of $S_0$ Lamb wave mode

c) Through thickness vector plot of $S_0$ Lamb wave mode

Figure 12. Through-thickness plots of (a) out-of-plane displacement, (b) in-plane displacement, and (c) vector field for $S_0$ Lamb wave mode at t=33.25 μs for fb/2=300 kHz-mm

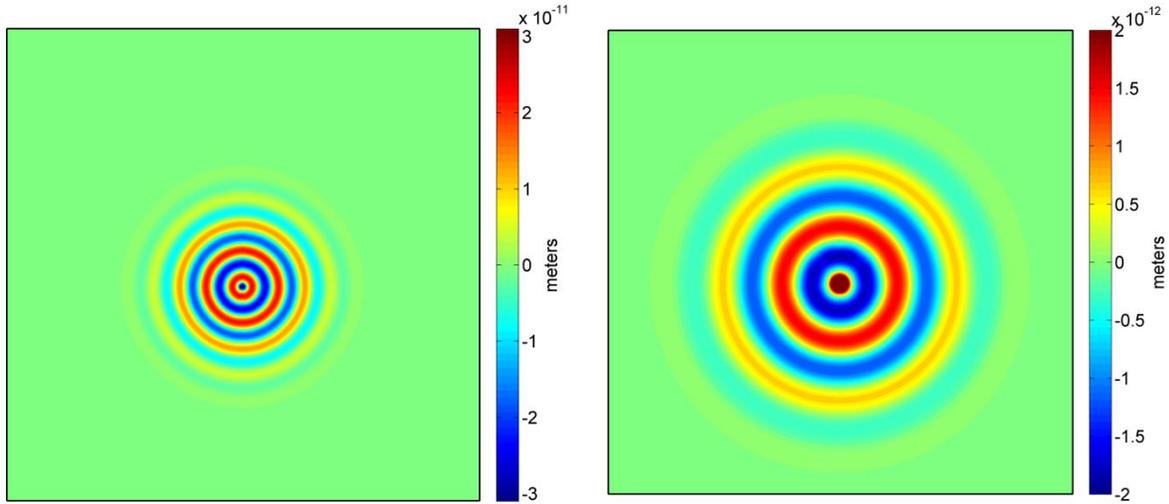

a) $A_0$ Lamb wave mode out-of-plane displacement  b) $S_0$ Lamb wave mode out-of-plane displacement
Figure 13. Out-of-plane displacement contour on the plate surface of the (a) $A_0$ and (b) $S_0$ Lamb wave modes at t=33.25 μs for fb/2=300 kHz-mm

**Imposing Stress-Free Boundary Condition**

In the past, when using LISA/SIM for Lamb wave analysis, researchers have imposed the necessary stress-free boundary conditions at the plate surfaces in one of two ways: 1) surrounding the plate with vacuum layers, Figure 14a, and 2) surrounding the plate with a combination of air and vacuum layers, Figure 14b. The vacuum layers were typically defined as having 1/10,000th of the stiffness of the plate material. The combination of air and vacuum layers is made up of a single layer of cells with the mechanical and material properties of air and the remainder of the surrounding layers defined as vacuum cells. A more physically accurate manner to impose the stress-free boundary condition on the plate is to surround it with multiple layers of cells with the mechanical properties of a fluid and the physical properties of air. Since no shear waves are able to propagate in fluids such as air, the stiffness matrix can be expressed, as seen in Equation (48), where K is the bulk modulus of air. The bulk modulus of a fluid can be expressed in terms of the density ($\rho$) and the speed of sound in the fluid ($v_{sound}$), as seen in Equation (49).

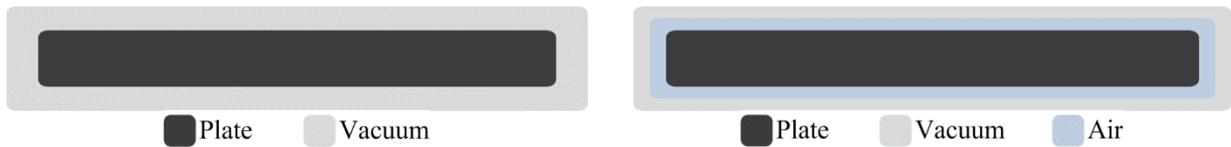

a) Plate surrounded with vacuum layers  b) Plate surrounded with one air and 'n' vacuum layers
Figure 14. Vacuum and air cells surrounding plate used to impose free-surface condition

$$C = \begin{bmatrix} K & K & K & 0 & 0 & 0 \\ K & K & K & 0 & 0 & 0 \\ K & K & K & 0 & 0 & 0 \\ 0 & 0 & 0 & 0 & 0 & 0 \\ 0 & 0 & 0 & 0 & 0 & 0 \\ 0 & 0 & 0 & 0 & 0 & 0 \end{bmatrix}$$  Equation (48)

$$K = \rho v_{sound}$$  Equation (49)

A study was conducted utilizing the developed model to test for convergence of the mean sensor signal as the number of surrounding layers was increased. If the necessary stress-free boundary condition is satisfied, the signal error should quickly converge to zero. Three convergence studies were conducted for surrounding layers with the properties of: 1) vacuum; 2) combination of air and vacuum; and 3) air with the mechanical properties of a fluid, and are presented in Figure 15a, Figure 15b, and Figure 15c, respectively. The plots present the mean signal error percentage as a function of the number of layers. The mean signal error is defined as the error between the sensor signal at the current number of layers and the sensor signal at the final number of layers investigated (i.e., ten total layers for Cases 1 and 2 and five layers for Case 3). Case 3 was only carried out to five layers because convergence was achieved.

In Figure 15a, it is evident that convergence is not reached as the number of vacuum layers increases. This is likely caused by numerical instability as a result of the physically inaccurate manner in which the free surface condition is imposed and the propagation and reflection of shear waves into the vacuum layers. Similarly, in Figure 15b, it is evident that convergence is not reached as the number of combined air and vacuum layers increases. This is also likely caused by numerical instability as a result of the physically inaccurate manner in which the free surface condition is imposed. However, in Figure 15c, it is evident that convergence is reached after approximately three layers. Even with a single layer, very little error in the sensor signal is present. Numerical instability was never found to be present with this approach, even after ten layers.

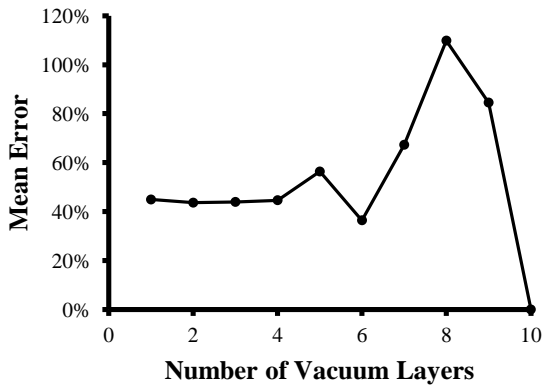
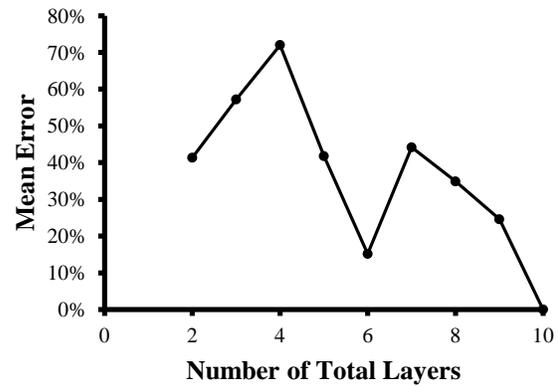

a) Mean signal error vs. number of vacuum layers surrounding plate

b) Mean signal error vs. number of air + vacuum layers surrounding plate

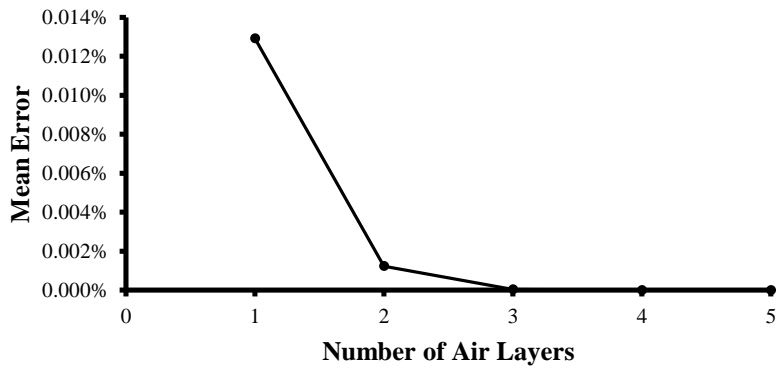

c) Mean signal error vs. number of air layers surrounding plate

Figure 15. Convergence of mean sensor signal using three kinds of boundary cells

By plotting the sensor signal from the simulations using air, vacuum, and air/vacuum cells surrounding the medium to enforce the free surface boundary condition, the numerical instability, denoted with the red oval, caused by the vacuum cells is evident. This instability is not obvious in the sensor signal of the first mode but becomes more pronounced as time progresses. Imposing the stress-free boundary condition in this way should be avoided since it does not provide a physically accurate means in which to model the boundary of the plate and because the numerical instability can cause significant error in the waveform following the arrival of the $S_0$ mode.

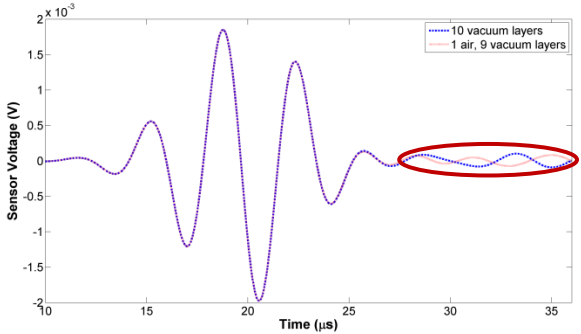
a) Sensor signal for ten air layers and ten vacuum layers

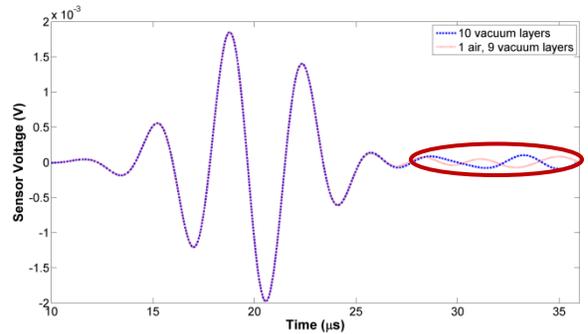
b) Sensor signal for ten air layers and 1 air + 9 vacuum layers

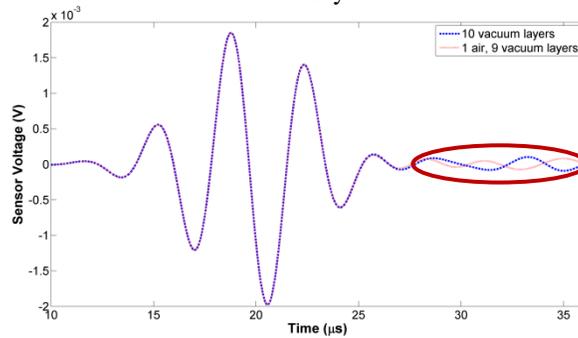
c) Sensor signal for ten vacuum layers and 1 air + 9 vacuum layers

Figure 16. Sensor signal comparison for three different boundary cells

**Computational Efficiency**

The LISA/SIM solution methodology, of which the current developed model is based, was formulated to run in a parallel processing environment. The computational efficiency of the current model offers key advantages over other wave propagation models. A 247 x 247 x 4 mm aluminum plate with one actuator and one sensor was modeled using the developed model and the commercial finite element software Abaqus (2007). Both models were run in a parallel computing environment on eight Harpertown 2.66 GHz, 8 MB/Cache, 16 GB memory processors. Each model was run in double precision for 1000 iterations with a time step of 9.5e-8 s. The computation results are shown in Table 3. Although the number of elements required for the current model (due to the surrounding air layers) was more than twice that required for the FEM model, the current model was significantly faster (>170 times) than the comparable FEM model. Using a time step of 9.5e-8, numerical instability occurred in the FEM model of the plate; a time step of 3e-8 s was required to resolve the issue of numerical instability. In addition, the FEM model under-predicted the theoretical wave speed by 13.3% while the result using the current model was within 4.1%.

Table 3. Computational efficiency comparison between FEM and current model

| Solver Method | # of elements | Wallclock time (s) |
|---|---|---|
| **FEM** | 244,038 | 40,855 |
| **Current Model** | 567,009 | 230 |

## CONCLUSION

A fully coupled electromechanical elastodynamic model for wave propagation in a heterogeneous, anisotropic material system was developed to investigate the physics of wave propagation, in particular Lamb wave propagation for the purpose of SHM. The model, derived using the LISA/SIM solution methodology, provides the capability of incorporating piezoelectric elements into a modeling scheme that has been previously proven to be a valuable tool for GW-based damage detection in isotropic and composite structures of arbitrary geometries and material architectures. The developed model was validated theoretically against the dispersion curve of an aluminum plate and proven capable of accurately simulating the group velocity of the $A_0$ and $S_0$ Lamb wave modes over a large range of frequency-thickness products. The through-thickness contour and velocity vector plots also verify the simulated Lamb wave out-of-plane and in-plane displacements match with the theoretical displacement profiles. Beside its accuracy in predicting wave speeds, the developed model was shown to be computationally efficient compared to finite element. Collocated actuators were modeled and the physics of Lamb wave mode suppression was investigated, including the relative energy of the modes as a function of frequency. The effect of actuation type was studied to determine the results from applying an equivalent displacement boundary condition on the actuator nodes to excite a GW instead of an electric potential across a piezoelectric element. It was found that inconsistent wave speed results occurred with displacement boundary condition actuation. A study comparing the piezoelectric sensor voltage to the displacement of the interface nodes was conducted and it was found that there exists a time lag between the in-plane nodal displacement and the sensor voltage. The developed model resulted in an accurate and efficient means to study the physics of GW propagation for SHM and assist in the development of SHM monitoring strategies.

## ACKNOWLEDGEMENTS

This work is supported in part by the National Science Foundation Graduate Research Fellowship under Grant No. (2011124478) and the MURI Program, Air Force Office of Scientific Research, Grant No. (FA9550-06-1-0309); Technical Monitor, Dr. David Stargel.

## REFERENCES

Abaqus Version 6.7, Abaqus/CAE and Abaqus/Explicit. 2007. Simulia World Headquarters, Providence.

Agostini, V., Delsanto, P. P., Genesio, I., and Olivero, D. 2003. "Simulation of Lamb Wave Propagation for the Characterization of Complex Structures," IEEE Trans. *Ultrason., Ferroelectr., Freq. Control*, Vol. 50, No. 4.

Alleyne, D. N., and Cawley, P. 1991. "A two-dimensional Fourier transform method for the measurement of propagating multimode signals," Journal of the Acoustical Society of America, Vol. 89, pp. 1159-1168.

Alleyne, D. N. and Cawley, P. 1992. "The Interaction of Lamb Waves with Defects," *J. IEEE Trans. Ultrason. Ferroelectr. Freq. Control,* Vol. 39, pp. 381–397.

Andrews, J. P., Palazotto, A. N., DeSimio, M. P., and Olson S. E. 2008. "Lamb Wave Propagation in Varying Isothermal Environments," *Structural Health Monitoring*, Vol. 7, pp. 265-270.

Balasubramanyam, R., Quinney, D., Challis, R. E., and Todd, C. P. D. 1996. "A finite-difference simulation of ultrasonic Lamb waves in metal sheets with experimental verification," Journal of Physics D-Applied Physics, Vol. 29, No. 1, pp. 147–155.

Chattopadhyay, A., Peralta, P., Papandreou-Suppappola, A., and Kovvali, N. 2009. "A multidisciplinary approach to structural health monitoring and damage prognosis of aerospace hotspots," *Journal of Royal Aeronautical Society*, Vol. 113, No. 1150, pp. 799-810.

Cheung, Y. K. 1976. *Finite Strip Method in Structural Analysis*, Pergamon, Oxford.

Cho, Y. and Rose, J. L. 1996. "A boundary element solution for mode conversion study of the edge reflection of Lamb waves," *J. Acoust. Soc. Am.,* 99:2097–2109.

Delsanto, P. P., Whitcombe, T., Chaskelis, H. H., and Mignogna, R. B. 1992. "Connection machine simulation of ultrasonics wave propagation in materials I: one-dimensional case," *Wave Motion*, 16:65–80.

Delsanto, P. P., Schechter, R. S., Chaskelis, H. H., Mignogna, R. B., and Kline, R. 1994. "Connection machine simulation of ultrasonics wave propagation in materials II: two-dimensional case," *Wave Motion*, 20:295–314.

Delsanto, P. P., Schechter, R. S., and Mignogna, R. B. 1997. "Connection machine simulation of ultrasonics wave propagation in materials III: three-dimensional case," *Wave Motion*, 26:329–39.

Diaz, S. and Soutis C. 2000. "Health monitoring of composites using Lamb waves generated by piezoelectric devices," *Plast. Rubber Compos.,* 29:475–481.

Ditri, J. and Rose, J. L. 1994. "Excitation of guided waves in generally anisotropic layers using finite sources," *J. Appl. Mech. (Trans. ASME),* 61:330–338.

Fornberg, B. 1998. *A Practical Guide to Pseudospectral Methods,* Cambridge University Press, Cambridge.

Giurgiutiu, V., Bao, J., and Zhao, W. 2003. "Piezoelectric Wafer Active Sensor Embedded Ultrasonics in Beams and Plates," *J. Exp. Mech.*, 43(4):428–449.

Giurgiutiu, V. 2005. "Tuned Lamb wave Excitation and Detection with Piezoelectric Wafer Sensors for Structural Health Monitoring," *J. Intel. Mat. Syst. and Str.*, 16:291-305.

Giurgiutiu, V. 2008. *Structural Health Monitoring with Piezoelectric Wafer Active Sensors*, Academic Press, Boston.

Guo, N. and Cawley, P. 1993. "The interaction of Lamb waves with delaminations in composite laminates." *J. Acoust. Soc. Am.,* 94:2240–2246.

Jha, R. and Watkins, R. 2009. "Lamb Wave Based Diagnostics of Composite Plates Using a Modified Time Reversal Method," AIAA-2009-2108, 17th AIAA/ASME/AHS Adaptive Structures Conference, Palm Springs, CA (May 4-7, 2009).

Koshiba, M., Karakida, S. and Suzuki, M. 1984. "Finite element analysis of Lamb waves scattering in an elastic plate waveguide," *IEEE Trans. Sonic Ultrason.,* 31:18–25.

Krawczuk, M. and Ostachowicz, W. 2001. "Spectral finite element and genetic algorithm for crack detection in cantilever rod," *Proc. 4th Int. Conf. on Damage Assessment of Structures (DAMAS) (Cardiff, UK, 25–28 June 2001)* (Switzerland: Trans Tech Publications) pp 241–50.

Lamb, H. 1917. "On waves in an elastic plate," *Proc. R. Soc. A*, 93:293–312.

Lee, B.C. and Staszewski, W. J. 2007. "Lamb wave propagation modelling for damage detection: II. Damage monitoring strategy," *Smart Mat. Struct.*, 16:260-274.

Liu, G. and Achenbach, J. D. 1995. "Strip element method to analyze wave scattering by cracks in anisotropic laminated plates," *J. Appl. Mech.,* 62:607–613.

Liu, G., Xi, Z., Lam, K. Y., and Shang, H. 1999. "A strip element method for analysing wave scattering by a crack in an immersed composite laminate," *J. Appl. Mech.,* 66:894.

Liu, Y., Mohanty, S., Chattopadhyay, A. 2010. "Condition Based Structural Health Monitoring and Prognosis of Composite Structures under Uniaxial and Biaxial Loading," *Journal of Nondestructive Evaluation*, Vol. 29, No. 3, pp. 181-188.

Liu, Y., Kim, S.B., Chattopadhyay, A., Doyle, D. 2011. "Application of System Identification Techniques to Health Monitoring of On-Orbit Satellite Boom Structures," *Journal of Spacecraft and Rockets*, Vol. 48, No. 4, pp. 589-598.

Moulin, E., Assaad, J., and Delebarre, C. 2000. "Modeling of Lamb waves generated by integrated transducers in composite plates using a coupled finite element–normal modes expansion method," *J. Acoust. Soc. Am.* 107:87–94.

Raghavan, A. and Cesnik, C. E. S. 2005. "Finite-dimensional piezoelectric transducer modeling for guided wave based structural health monitoring," *Smart Mater. Struct.*, 14:1448-1461.

Raghavan, A. and Cesnik, C. E. S. 2007. "Review of Guided-wave Structural Health Monitoring," *Shock Vib. Dig.*, 39(2):91-114.

Rose, L. R. F. and Wang, C. H. 2004. "Mindlin plate theory for damage detection: source solutions," *J. Acoust. Soc. Am.* 116:154–171.

Santosa, F. and Pao, Y. H. 1989. "Transient axially asymmetric response of an elastic plate," *Wave Motion II,* 11:271–295.

Sohn, H. and Kim, S. B. 2010. "Development of Dual PZT Transducers for Reference-Free Crack Detection in Thin Plate Structures," IEEE Trans. *Ultrason., Ferroelectr., Freq. Control*, 57(1):229-240.

Soni, S., Das S., and Chattopadhyay, A. 2009. "Simulation of damage features in a lug joint using guided waves," *Journal of Intelligent Material Systems and Structures*, Vol. 20, No. 12, pp. 1451–1464.

Staszewski, W. J., Pierce, S. G., Worden, K., Philp, W. R., Tomlinson, G. R., and Culshaw, B. 1997. "Wavelet Signal Processing for Enhanced Lamb-Wave Defect Detection in Composite Plates Using Optical Fiber Detection," *J. Opt. Eng.*, 36:1877–1888.

Sundararaman, S. 2007. "Numerical and Experimental Investigations of Practical Issues in the use of Wave Propagation for Damage Identification," Ph.D. Dissertation, Department of Mechanical Engineering, Purdue University.

Talbot, R. J. and Przemieniecki, J. S. 1976. "Finite element analysis of frequency spectra for elastic wave guides," *Int. J. Solids Struct.,* 11:115–138.

Veidt, M., Liu, T., and Kitipornchai, S. 2001. "Flexural waves transmitted by rectangular piezoceramic transducers," *Smart Mater. Struct.* 10:681–688.

Viktorov, I. A. 1967. *Rayleigh and Lamb Waves,* Plenum, New York.

Virieux, J. 1986. "P-SV wave propagation in heterogeneous media: Velocity stress finite difference method," Geophysics, Vol. 51, No.4, pp. 889-901.

Wilcox, P. 2004. "Modeling the excitation of Lamb and SH waves by point and line sources," *Review of Quantitative Nondestructive Evaluation* vol 23, ed D O Thompson and D E Chimenti pp 206–13.

Yamawaki, H. and Saito, T. 1992. "Numerical calculation of surface waves using new nodal equation," *Nondestruct. Testing Evaluation,* 8/9:379–389.

Zienkiewicz, O. C. 1989. *The Finite Element Method* 4th edn., McGraw-Hill, New York.